\documentclass[english,a4paper,12pt]{article}
\usepackage[T1]{fontenc}
\usepackage{babel}
\usepackage{graphicx}
\usepackage{epsf}
\usepackage{cite}
\usepackage[dvips]{color}
\begin{document}
\noindent
\begin{center}
{\bf \Large
Cubic boron nitride: a new prospective material for ultracold
neutron application
}
\end{center}
\normalsize
\noindent
Yu.\ Sobolev$^{1,a,*}$, Th.\ Lauer$^2$, Yu.\
Borisov$^3$, M.\ Daum$^4$, N.\ du Fresne$^1$, L.\ G\"oltl$^{2,8}$, G.\ Hampel$^2$,
W.\ Heil$^1$, A.\ Knecht$^{4,5}$, M.\ Keunecke$^6$, J.\ V.\ Kratz$^2$, T.\ Lang$^1$,
M.\ Meister$^1$, Ch.\ Plonka-Spehr$^2$,
Yu.\ Pokotilovski$^7$, P.\ Reichert$^2$, U.\ Schmidt$^8$,\ Th.\ Krist$^9$,
N.\ Wiehl$^2$,\ J.\ Zenner$^1$ \\

\noindent
$^1$Institute for Physics, Johannes Gutenberg-Universität Mainz, Staudingerweg\ 7, \\
D-55128 Mainz, Germany\\
$^2$Institute for Nuclear Chemistry, Johannes Gutenberg-Universit\"at Mainz,\\
Fritz-Strassmann-Weg 2, D-55128 Mainz, Germany \\
$^3$Petersburg Nuclear Physics Institute, 188300 Gatchina, Leningrad region, Russia \\
$^4$PSI, Paul Scherrer Institut, CH 5232 Villigen, Switzerland\\
$^5$University of Zurich, Switzerland\\
$^6$Fraunhofer-Institut f\"ur Schicht- und Oberfl\"achentechnik, Bienroderweg 54e,\\
D-38108 Braunschweig, Germany \\
$^7$JINR, Joint Institute for Nuclear Research, 141980 Dubna, Moscow region, Russia\\
$^8$Physical Institute, Universit\"at Heidelberg, Philosophenweg 12,\\
D-69120 Heidelberg, Germany \\
$^9$Hahn-Meitner-Institut Berlin, Glienicker Str. 100, D-14109 Berlin, Germany

\begin{center}
{\bf Abstract}
\end{center}
\noindent
For the first time, the neutron optical wall-potential
of natural cubic boron nitride (cBN) was measured at the ultracold neutron (UCN) source
of the research reactor TRIGA Mainz using the time-of-flight method (TOF).
The samples investigated had a wall-potential of ($305 \pm 15$) neV.
This value is in good agreement with the result
extracted from neutron reflectometry data and
theoretical expectations.
Because of its high critical velocity for UCN and
its good dielectric characteristics, cubic boron nitride  coatings (isotopically
enriched) will be useful for a number of applications in UCN experiments.\\

\noindent
$^*$Corresponding author; email: iouri@uni-mainz.de\\
$^a$On leave from Petersburg Nuclear Physics Institute, Gatchina, Russia\\

\noindent
{\bf PACS numbers and key words:}\\
28.20.Gd, Neutron transport: diffusion and moderation\\
14.20.Dh, Protons and neutrons \\
29.25.Dz, Neutron sources

\newpage
\noindent
{\bf \Large 1. Introduction}\\
Ultracold neutrons can be stored in material bottles for
times approaching the beta decay lifetime of the neutron. The
storage is based on the reflection of the UCN by selected materials
under any angle of incidence. The reflection is caused by the
coherent strong interaction of the neutron with the atomic nuclei.
Quantum mechanically, this can be described by an effective
potential which is commonly referred to as Fermi pseudo potential
or the material optical potential, $U_\mathrm{f}$.

The storage of neutrons
with very low energies was predicted by Zeldovich \cite{zeldovich}
in 1959 and experimentally realized simultaneously by groups in
Dubna \cite{lushikov} and in Munich \cite{Steyerl} in 1968. Since
then, UCN have become a unique tool for fundamental neutron physics,
e.g., experiments looking for a permanent electric dipole moment
of the neutron (nEDM) \cite{nedm_ill} or measuring the neutron
lifetime \cite{morozov,serebrov}. UCN have velocities typically below 7 m/s,
corresponding to energies below 250\,neV.

The material
optical potential $U_\mathrm{f}$ can be expressed as $U_\mathrm{f} = V - \imath W$
\cite{Fermi}, where the wall-potential $V$ describes the reflecting and $W$ the
absorbing behaviour of the material and is given by
\begin{equation}
\label{eqn:Potentials}
V=\frac{2\,\pi\,\hbar^2}{m}\cdot\sum_i N_i\, b_i,\quad\quad
W=\frac{\hbar}{2}\cdot\sum_i N_i \, \sigma_i\,v_\mathrm{th}.
\end{equation}
Here, $m$ is the neutron mass, $N$ is the scattering center
density, $b$ is the bound coherent scattering length, $\sigma$ is
the sum of the absorption, $\sigma_\mathrm{a}$, and the inelastic scattering cross
section, $\sigma_\mathrm{ins}$, at thermal neutron velocities, $v_\mathrm{th}$. The sum over
$i$ includes all specific nuclei of the chemical composition.
Neutrons with a normal velocity component below a critical value
$v_\mathrm{c} = (2\,V/m)^{1/2}$ are totally reflected from the surface. The
loss probability per reflection is expressed by the
energy-independent loss coefficient $\eta={W}/{V}$.

The $v^2$ dependence of the UCN velocity spectrum leads to a
storable UCN density which scales with $V^{3/2}$. Therefore, $V$ has
to be as high as possible to increase statistics in UCN
experiments. At the same time, the $\eta$ value has to be low
enough to reduce absorption and upscattering losses. Besides this, specific
experiments like the nEDM require additional properties of the
materials, e.g., high electrical resistivity (also against surface
currents), non-magnetic properties, and high vacuum
compatiblility.

The material with the so far highest known wall-potential commonly used in
UCN experiments, is $^{58}$Ni with $V$ = 346 neV
corresponding to a critical velocity of $v_\mathrm{c} = 8.1\,\rm m/s$.
Nickel, however, is magnetic and conductive. So far, only
beryllium oxide (BeO, $V$ = 261 neV, $\eta < 1\times 10^{-4}$)
satisfies all criteria mentioned above. But its toxicity may be
prohibitive.

Looking for other non-toxic, non-magnetic materials with high wall
potentials and high electrical resistivity, we got interested in
cubic boron nitride (cBN). Due to its hardness and high wear
resistance, cBN is under development as a super-hard coating
(micro-hardness $>\,\rm 50\, GPa$, \cite{Keunecke}) for cutting tools in the
metal machining industry.

Natural boron consists of two isotopes of abundance ratio: $\rm
^{10}B/^{11}B=0.248$. The calculation of the pseudo-potential of
this chemical composite using
Eq.\ (\ref{eqn:Potentials}) and
the given material parameters, $\rho\approx 3.5\,\rm g/cm^3\rm $ \cite{Jiang, Lehmann},
$N \approx$ 8.5$\times 10^{22}\,\rm cm^{-3}$,
$b(^{10}\rm B)=-0.1\,fm$, $b(^{11}\rm B)=6.65\,fm$, $b$(N)=9.36 fm, yields
$V$ = 324\,\rm neV and $\eta \approx 1.5\times 10^{-2}$. The value for
$\eta$ reflects the enormously high absorption cross section of
$^{10}$B, $\sigma_{a}=3840$ barn\cite{NIST}.

This drawback can, in
principle, be overcome by using isotopically enriched $^{11}$B in the composite
($\sigma_{a}=0.0055$ barn), which results in an even higher
wall-potential of $V$ = 351\,neV. The value of $\eta \approx 3.3\times
10^{-5}$ in this case is mainly dominated by the absorption of
$^{14}$N ($\sigma_\mathrm{a} = 1.9$ barn) for an $^{11}$B enrichment of > 99.95 \%.
As a first step, we started with the investigations of natural cBN.\\

\noindent
{\bf \Large 2. Experiments}\\
We carried out two experiments to measure the critical velocity of UCN reflected from 
natural cBN using (i) a time-of-flight (TOF) measurement in order to determine
the transmission of ultracold and very cold neutrons (VCN)
through a silicon wafer (380 $\mu$m) coated with natural
cBN of thickness 300 nm and (ii) cold neutron reflectometry with the
same sample. For comparison, a natural nickel coated sample 
with a layer thickness of 500 nm was investigated using the TOF method.

The cBN films were prepared in a reactive r.f.\ diode sputtering
system using an electrically conducting boron carbide (B$_4$C)
target. In order to reach good adhering cBN layers, a special
process was used starting with an interlayer of boron carbide and
continuing with graded BCN interlayers, where the carbon is replaced by nitrogen
by means of an incremental change from argon to
nitrogen as sputter gas. Further information on the process and
the sputtering facility can be found in Refs.\ \cite{Keunecke,Yamamoto,Yamamoto2}.
With this method, cBN layers
of up to 2 $\mu$m thickness and a cubic phase content of
approximately 90 \% (measured by IR spectroscopy
\cite{Keunecke,Yamamoto,Yamamoto2}), corresponding to a density of
3.3 g/cm$^3$, are produced.\\
The Nickel coating was produced at Hahn-Meitner-Institut (HMI) by conventional DC magnetron sputtering.
With this technique it is possible to reach coating densities close to 100$\%$ of bulk material.\\

\noindent
{\bf \large 2.1 The time-of-flight experiment}\\
In the time-of-flight (TOF) experiment,
we determined the neutron optical potential via the
critical neutron velocity from the transmission of slow neutrons
through the sample.
The experiment was performed at the solid deuterium
UCN source at beamport C \cite{AFrei} of the research
reactor TRIGA Mainz\cite{eberhardt,hampel}.
This reactor can be operated at a steady-state power of 100 kW or in pulsed mode with a
maximum power of 250 MW and an energy release of 10 MWs\cite{menke}.

In principle, a TOF measurement can be performed both in the pulsed or in
steady state mode of the reactor. In the pulsed mode, the minimal overall length of the flight path
is from the layout of the reactor (biological shield, cryostat,
etc., cf.\ \cite{AFrei}) more than 6 m.
In such a long UCN guide, part of the neutrons are repeatedly diffusely scattered back and forth
and the neutron guide acts also as a neutron storage chamber.
This leads to a delay of these scattered neutrons in reaching the detector
apparently at a lower velocity which influences the results.
In the steady state mode using
a chopper for the TOF information, the TOF path length can be chosen to be
much shorter and the delay effect is negligible.

\begin{figure}[t]

\begin{center}
\includegraphics[width=110mm]{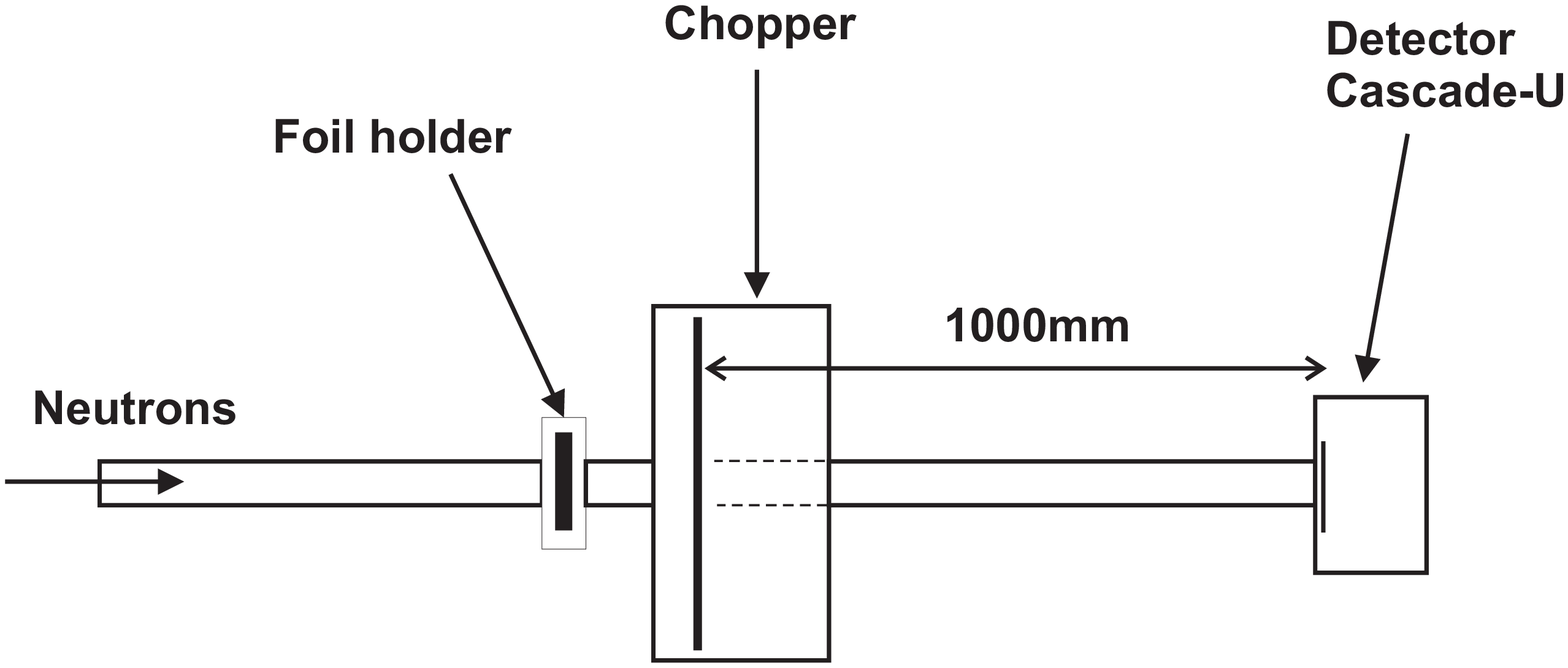}

\end{center}

\caption{~}{Sketch of the experimental setup for the time-of-flight experiment with
a chopper.
}

\label{fig_chopper}

\vspace{6cm}

\end{figure}

Therefore in this experiment, the reactor was operated in the steady-state mode of 100 kW thermal power.
In all measurements, the same amount of solid deuterium (4 mol) in
the UCN source was used.
Outside the biological shield
of the reactor core, at a distance of $\sim$4 m from the solid $^2$H$_2$ converter,
the in-pile part of the neutron guide is terminated by an aluminium window at room temperature.
We used a time-of-flight spectrometer with a chopper \cite{Fierlinger} similar to that used in
Refs.\ \cite{Altarev,Atchison,Atchison1}.
The neutrons were guided via electro-polished stainless steel tubes with inner
diameter 66 mm from the UCN source to the TOF spectrometer.
The arrangement is schematically sketched in Fig.\,\ref{fig_chopper}.
Neutrons enter from the left and pass the TOF spectrometer towards
a Cascade-U detector \cite{klein} at the end of the flight path.
The flight path between chopper and detector was ($1006 \pm 3$)mm.
The Al entrance window of the detector was 0.1 mm thick with an energy barrier of 54 neV from the
material optical potential of aluminium and caused some
reduction of the detector efficiency.
With this setup, the (averaged) velocity component
along the forward direction can be measured.

The coated silicon wafers (diameter 76 mm) were inserted in a foil
holder placed in front of the chopper operating at a duty cycle of 4.0\,\% and
at a frequency of 1.0 Hz.
For the correct determination of a velocity distribution with the TOF method,
we assume a high guide transmission for UCN, {\it i.e.}\ a high
percentage of specular reflections in the UCN guide system,
which is experimentally supported by Ref.\cite{plonka}.
Without sample, the count rate in our TOF spectrometer was about 45 s$^{-1}$.

\begin{figure}[t]

\begin{center}
\includegraphics[width=110mm]{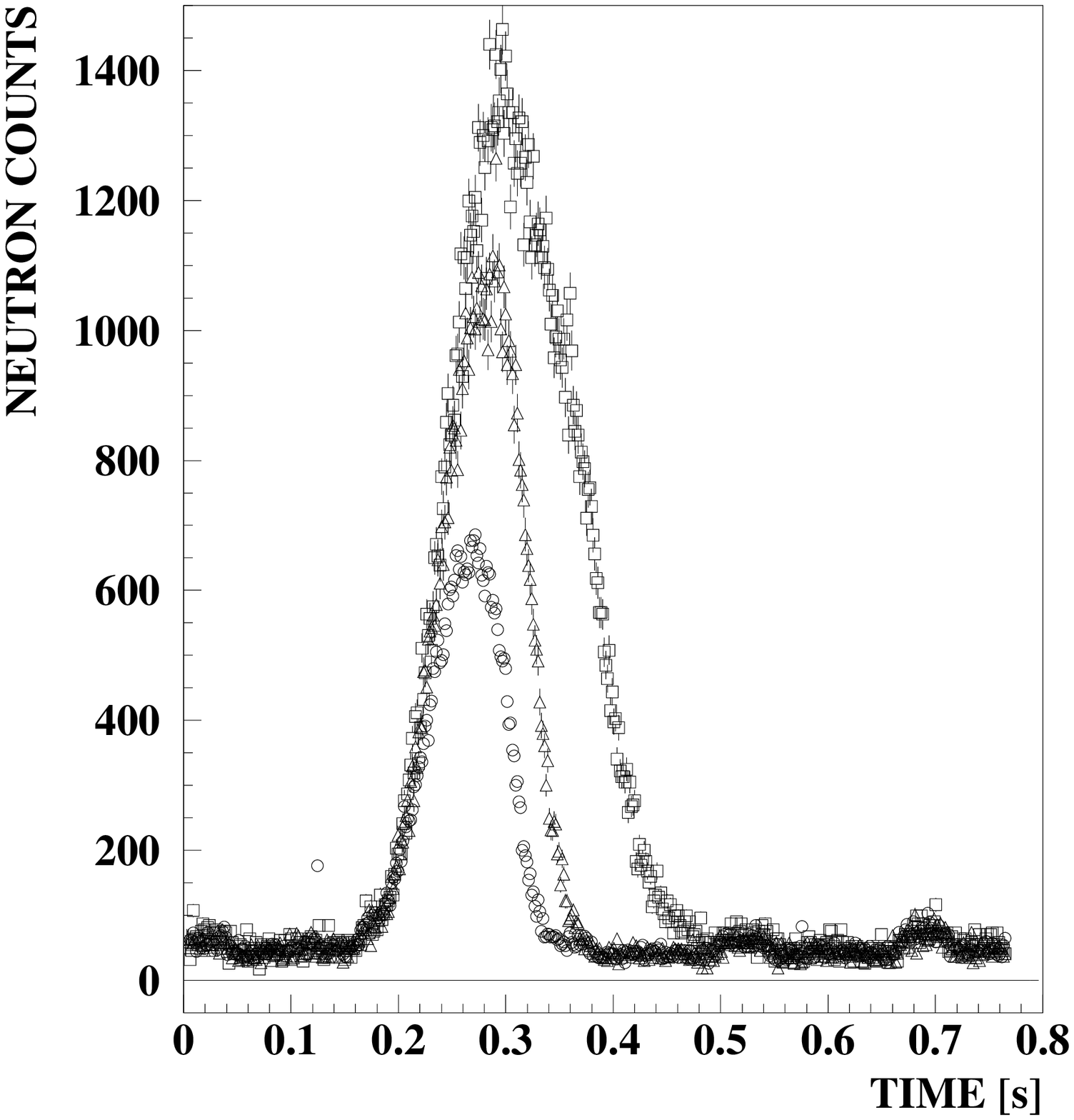}

\end{center}

\caption{~}{
Measured time-of-flight data (i) without sample (quads),
(ii) with the nickel coated silicon wafer (triangles) and the cBN coated one
(circles).
}

\label{fig_org}

\vspace{6cm}

\end{figure}

As DAQ system, we used a standard TOF card from the FAST company. The TOF spectrum
was started with the pulse from the chopper indicating the chopper status 'open'.
Neutrons were registered with a dwell time of 1.5 ms
according to their arrival time in the detector.
Figure \ref{fig_org} shows the time-of-flight spectra
without and with samples, {\it i.e.}, the nickel and cubic natural-boron nitride sample.\\

\noindent
{\bf 2.1.1 Background subtraction}\\
The data in Fig.\ \ref{fig_org} show backgrounds originating
mainly from three effects:
(i) a constant background which originates from thermal neutrons which are present
inside the experimental area during reactor operation and cannot be suppressed completely;
(ii) an almost constant background from neutrons which were diffusely scattered,
{\it i.e.}, the neutron guide acts also as a UCN storage chamber which is repeatedly filled
by the chopper cycles and emptied through the opening to the detector;
(iii) the periodic humps in the
spectra stem from very cold neutrons which could penetrate thinner parts
of the rotating chopper plates. These were to compensate for the openings in the plates
in order to avoid unbalances (they are now painted with the high absorption material
Gd$_2$O$_3$).

\begin{figure}[t]

\begin{center}
\includegraphics[width=110mm]{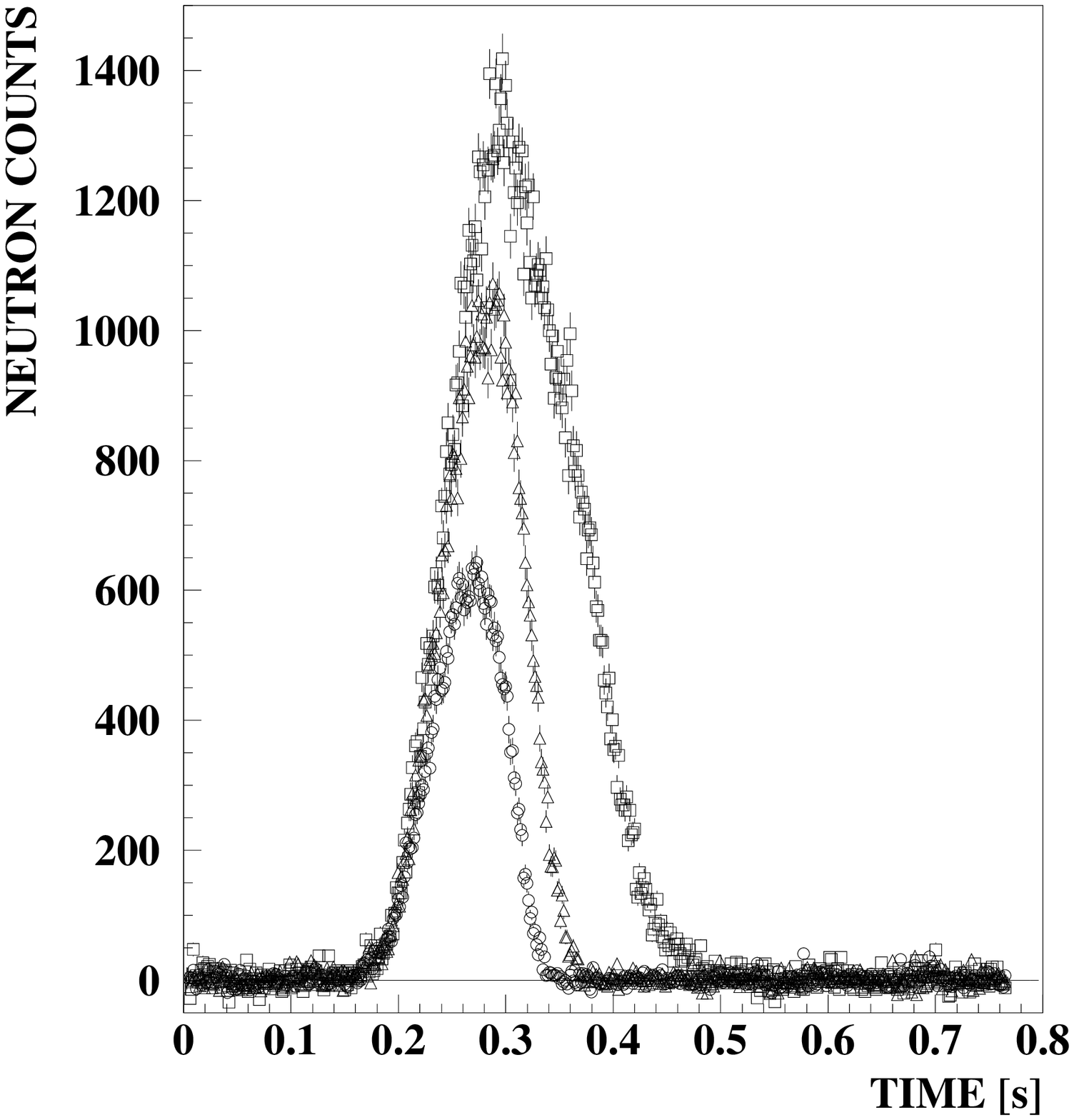}

\end{center}

\caption{~}{
Measured time-of-flight data after background subtraction
(i) without sample (quads),
(ii) with the nickel coated silicon wafer (triangles) and the cBN coated one
(circles).
}

\label{fig_cor}

\vspace{6cm}

\end{figure}

We divided the time scale of the three spectra into 10 regions
and fitted the background to constant values in seven regions left and right of
the TOF peaks. The background underneath the peaks was subtracted by
interpolation. The reduced $\chi^2$ of the fits ranged from 1.04 to 3.1 with one as high as 5.7.
The $\chi^2$ values, partly significantly larger than 1,
originate mainly
from sparks in the detector during the first days of operation.
These sparks were produced by short high voltage break throughs
from the gas electron multiplier (GEM) to the read-out structure of the detector
and led to an increased number of events in several channels, see, e.g.,
at t = 0.12 s in the cBN data in Fig.\ \ref{fig_org}.
Surface contaminations (dust particles)
on the GEM foil lead to a locally increased electric field
and sparks which also clean the surface.
Consequently, these contaminations were reduced
and the detector was more stable after several days. In order to take these
non-statistical uncertainties into account, we firstly replaced the very few
data which differed by more than 5 standard deviations from their
neighbours by the average of the neighbouring data. Secondly, we increased
the uncertainties of the respective data
points by a factor $\sqrt{\chi^2}$, as recommended, {\it e.g.}, by Refs.\cite{PDG1,PDG2}.
The data after background subtraction used for the further analysis
are shown in Fig.\ \ref{fig_cor}.\\

\noindent
{\bf 2.1.2 Time calibration}\\
Time calibration of the system, see Ref.\ \cite{Altarev}, was obtained by comparing
(i) at one distance the spectra taken at two different chopper frequencies;
(ii) at one chopper frequency the TOF spectra taken at the two distances.
We obtained eight partially dependent
calibration measurements by comparing the TOF spectra at different chopper frequencies,
three pairs at each distance: (a)1.00 Hz, 0.75 Hz, (b)1.00 Hz, 0.50 Hz, and (c) 0.75 Hz, 0.50 Hz,
and additional three calibration measurements (one per frequency)
by comparing the TOF spectra at the two distances.

Since the electronic signal from the chopper is produced by a photoelectric barrier triggered by an
interrupter on one of the rotating chopper discs, we obtain a
time difference, $\delta t$, between the electronic trigger signal from the
chopper and its real opening
time. The value of $\delta t$ depends on the chopper frequency.
Averaging all available and independent information
we find $\delta t(\mathrm{1 Hz}) = (0.191 \pm 0.003)$ s.
Recently, the calibration measurements were repeated at ILL with frequencies
0.5 Hz, 1 Hz, and 1.25 Hz confirming the earlier result \cite{Altarev}.\\

\noindent
{\bf \large 2.2 Analysis of the time-of-flight experiment}\\
In order to extract the values of the optical potential from the layers of interest, 
we developed an analysis procedure following three steps. 
In the first step, we approximated the measured data with
no sample and after background subtraction
with a cubic spline fuction based on nine data points
in order to obtain an analytic expression for the data.
The $\chi^2$ per degree-of-freedom was 2.8 which
on the one hand originates from sparks in the detector and on the other hand
reflects the rather simplified parameterization of the spectrum
by the spline which, however, is sufficient for our purposes.
In order to take this
non-statistical uncertainty into account, we increased the uncertainties of the data
points by a factor $\sqrt{\chi^2}$, see above.

In a second step, we calculate the transmission of the UCNs through layers.
For given values of the Fermi potential of each layer, we developed
procedures to perform a one-dimensional numerical solution of the Schrödinger equation.
Hereby, we include surface roughness via smearing of the potential step at the
border of adjacent layers. Matching the numerical solution of the
Schrödinger equation outside the layers with the well known analytical
solution for plane wave propagation allows us to calculate the value
for the transmission.

\begin{figure}[t]

\begin{center}
\includegraphics[width=110mm]{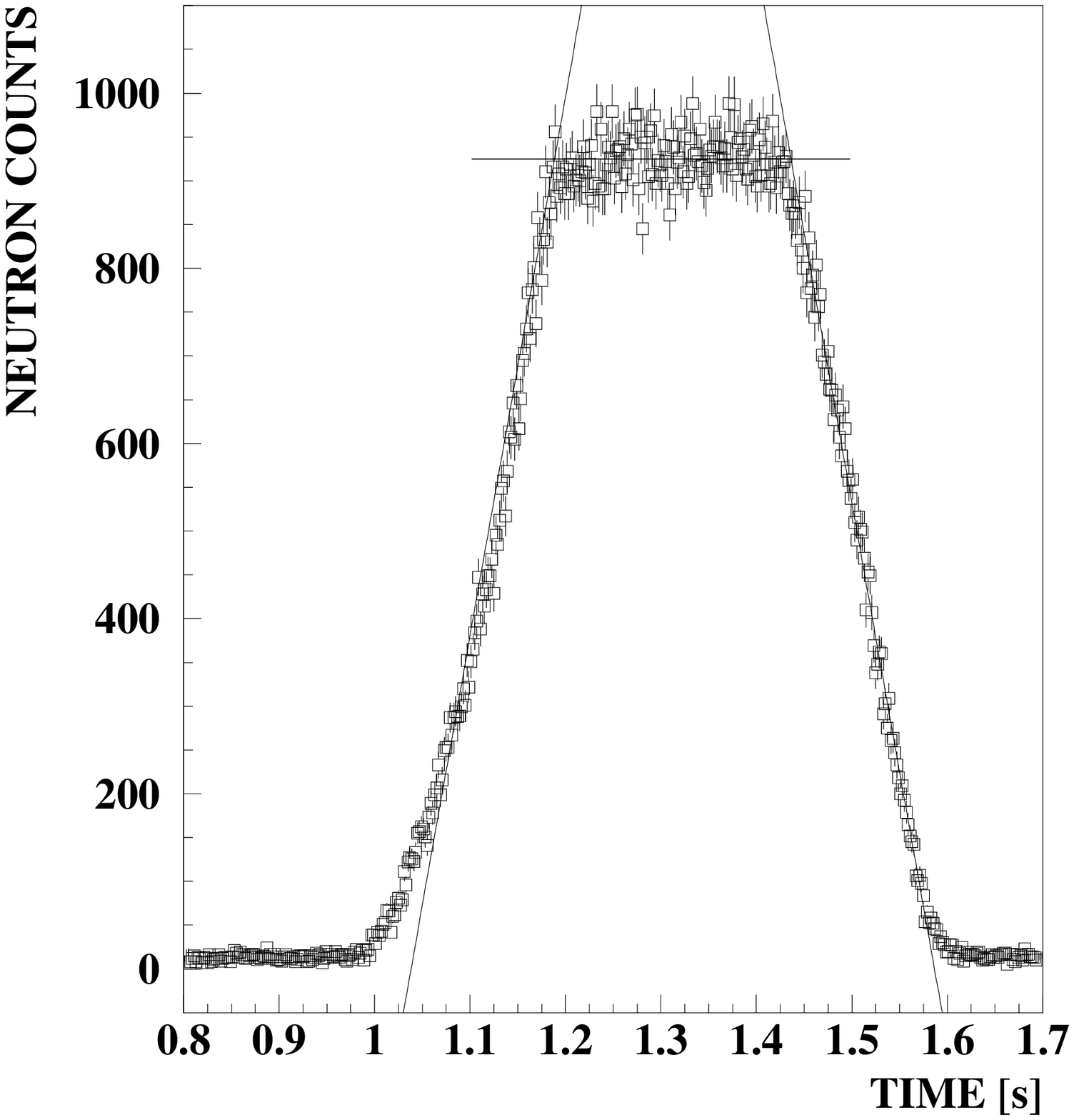}

\end{center}

\caption{~}{
Opening function of the chopper measured at a frequency of 0.1 Hz.
In the analysis, the opening function is approximated by the
trapezium as indicated.
}

\label{fig_trapez}

\vspace{6cm}

\end{figure}

\begin{figure}[t]

\begin{center}


\includegraphics[width=85mm]{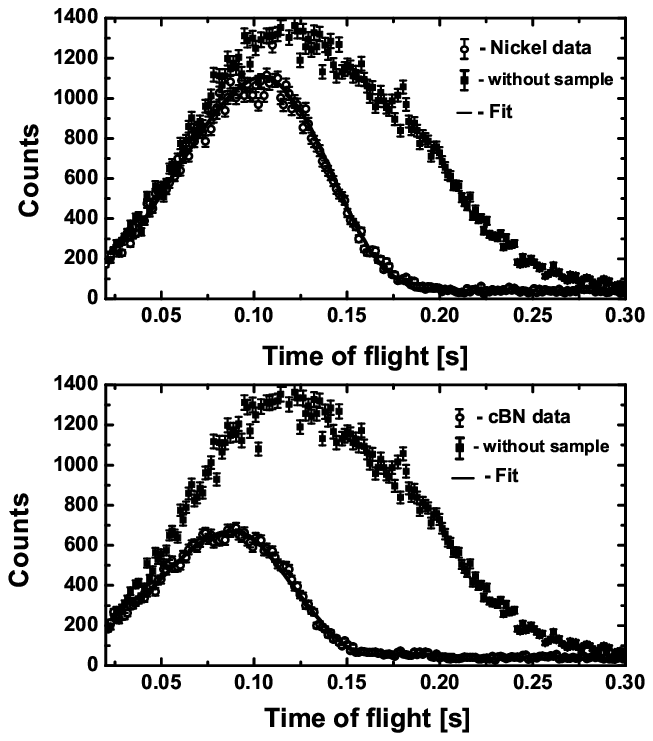}
\end{center}

\vspace{-1cm}

\caption{~}{
Time-of-flight data of very slow neutrons
through coated silicon wafers and the fit
to the data.
Top: 500 nm nickel coated on a 380 $\mu$m silicon wafer.
Bottom: 350 nm cBN on a 380 $\mu$m silicon wafer.
The corresponding TOF-data without sample
used to calculate the spline fitting function for unfolding the data
with the time acceptance of the chopper are shown in both plots.
The measured data
were corrected for the chopper time offset $\delta$t (see text).
}

\label{fig_5}

\vspace{6cm}

\end{figure}

\begin{figure}[t]

\begin{center}
\includegraphics[width=110mm]{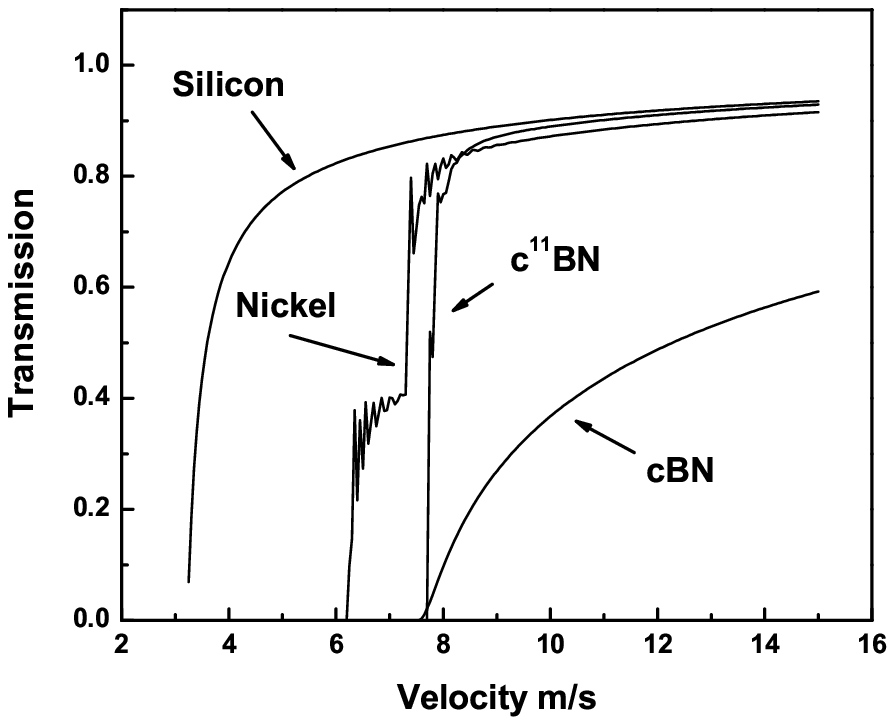}

\end{center}

\caption{~}{
Calculated transmission of the measured silicon wafers coated with Ni (500 nm) and cBN (300 nm) layers.
The calculation was performed by using the obtained fit parameters.
The theoretical transmission of an isotopically enriched c$^{11}$BN (300nm) layer was calculated
from literature cross sections. The saw-shape behaviour of transmission curves in some
places is the result of quantum interference at thin layers.}

\label{fig_6}

\vspace{6cm}

\end{figure}

In the third step, we compare our model with the measured data. 
For this purpose, we convolute the product of the transmission function (step 2)
and the initial spectrum (step 1) with the opening function of the chopper, 
see Fig.\ \ref{fig_trapez}. The free parameters of the fit were the
mean free paths in the coated layers (cBN, Ni), the loss parameter $\eta$
and the roughness of the sample. The values of these parameters were obtained 
by comparison of our model with the data and minimisation using 
standard least square fitting techniques.
The experimental data with the final fit are shown in Fig.\ \ref{fig_5}. 
In Fig.\ \ref{fig_6}, the curves for the velocity dependent transmission
for the different measured samples as derived from the analysis are shown.

Due to its ferromagnetic properties, the neutron transmission through the 
Ni sample results in the sum of the transmission for the different neutron spin states
(with their different optical potential values). From the measured data,
we extract the value of the optical potential for spin down of (207 $\pm$ 2) neV
and for spin up (284 $\pm$ 2) neV (here $\pm$ 2 neV includes uncertainty of fit only). 
The reduced $\chi^2$ of the fit is 4.5, see above.
This results correspond to a Ni bulk density of 98 \%
of the nickel density.

For our cBN sample, we obtain a value for the optical potential of (305 $\pm$ 2) neV.
From this, we calculate a value of 90 \% for the bulk density. 
The reduced $\chi^2$ of the fit was 1.7.

In order to take these $\chi^2$ values as non-statistical uncertainties into account,
we increase the uncertainties ($\pm$ 2 neV) by the factors $\sqrt{\chi^2}$.
The main contributions to the total uncertainty arise, however, from the time calibration
of the TOF spectrometer, $\delta V$ = 15 neV; the uncertainty originating from the 
flight path uncertainty is 2 neV. Thus, the total uncetainties of our 
values for the optical potentials are $\Delta V \approx $ 15 neV. The final results of our analysis
are summarized in Table I. While the optical potential of cBN has never been measured before,
the values for Ni are in excellent agreement with literature values\cite{Atchison}.

\noindent
{\bf \large 2.3 The neutron reflectometry experiment}\\
The second experiment using neutron reflectometry was performed
at the neutron reflectometer V14 at the cold beam at the
HMI. This method measures the critical
angle for total reflection of cold neutrons at grazing incidence,
where the velocity component normal to the surface is
comparable to the velocity of UCN. Unlike UCN, cold neutrons do not suffer from
small angle scattering inside the sample. Due to this fact, cold neutron reflectometry
can be seen as cross-cheque to the transmission measurements.

The neutron critical angle of total reflection, $\theta_c$, is
directly connected with the scattering length density, {\it i.e.}, the
term $N\cdot b$ in equation (\ref{eqn:Potentials}), via the
formula
\begin{equation}
\frac{\sin \theta_c}{\lambda}  = \sqrt{\frac{N \cdot b}{\pi}}
\label{iouri1}
\end{equation}
where $\lambda$ is the neutron wavelength (4.9 \AA\,\rm at V14). By
measuring $\theta_c$ for the samples under investigation, $V$ can
be determined using Eq.\ref{eqn:Potentials}:
\begin{equation}
V = \frac{2 \pi^2 \hbar^2 \sin^2 \theta_c}{m \cdot \lambda^2}
\end{equation}
The reflectivity function for the analysis of the data was derived with the same 
quantum mechanical calculation as mentioned above. However, because of the divergent
beam profile, the obtained reflectivity function has to be convoluted with the known
neutron beam profile.

\begin{figure}[t]

\begin{center}
\includegraphics[width=110mm]{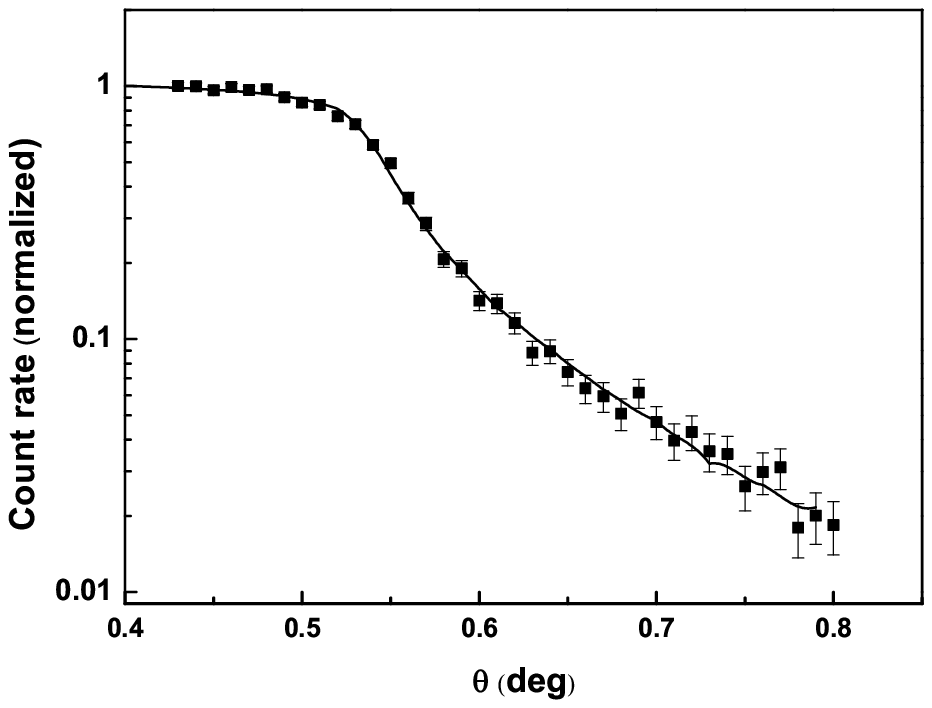}

\end{center}

\caption{~}{
Measured cold neutron reflectometry data from a silicon wafer coated with 300 nm of
cBN and the corresponding fit.
}
\vspace{5cm}

\label{fig_7}

\end{figure}

In the fit, the free parameters were
(i) the scattering length density $N \cdot b$, cf. Eq.\,\ref{iouri1},
and (ii) the roughness and
composition of the surfaces (substrate and coating).
The model of the cBN sample is represented by three layers, air, cBN, and silicon.
The scattering lengths, $b$, for air, boron, nitrogen, and silicon were taken from
the literature\cite{sears}.
The reduced $\chi^2$ of the fit to the reflectivity data is 3.0.
The uncertainties of cold neutron reflectometry depend strongly on the
flatness of the samples under investigation. Silicon wafers usually have
very smooth surfaces but not necessarily a high flatness over the whole
coated sample. This increases the uncertainty of the result.
In order to take such an (unknown) apparative uncertainty into account,
the experimental uncertainties were scaled\cite{PDG1,PDG2}
to obtain a $\chi^2$ of one.
Figure \ref{fig_7} shows the experimental data and
the fit. The extracted value of $V$ for the reflectometer data is
shown in Table \ref{results}.\\

\noindent
{\bf \Large 3. Conclusions and outlook}\\
The optical potential value $V$ of natural cubic boron nitride was determined experimentally
for the first time via the energy dependent transmission of very slow neutrons through a 
cBN layer, $V_\mathrm{trans} = (305 \pm 15)$ neV and by cold neutron reflectrometry,
$V_\mathrm{refl} = (300 \pm 30)$ neV. Both values agree with each other and are in excellent 
agreement with theoretical expectations, $V$ = 308 neV, obtained from Eq.\ \ref{eqn:Potentials} 
and for the fraction of the cubic phase, which was determined independently
by infrared spectroscopy to be about 90 \% \cite{Keunecke,Jiang,Lehmann}.
Using isotopically enriched c$^{11}$BN samples with a fraction of $^{11}$B > 99.95 \%, 
we expect a 10 \% increase of $V$ and a significantly reduced loss value $\eta$, 
as shown in Fig.\ \ref{fig_6}. This will be demonstrated in the near future.

In addition to the experimental determinations of the optical potential,
the insulator properties of the cBN samples were investigated
using a high resistance meter 4339 A in combination with the resistivity cell
16008B (Hewlett Packard). We found values for the specific resistivity 
R > 3.4$\cdot 10^{16}$ ohm$\cdot$cm. This is comparable to other high resistivity
values, {\it e.g.} of natural diamond or quartz.

In conclusion, cubic boron nitride is a very promising material for coatings 
with high optical potential and high resistivity. 

Furthermore, it is certainly worth while to investigate cubic carbon nitride, 
cC$_3$N$_4$, and isotopically pure $^{11}$B$_4$C as further coating materials 
for the physics with ultra-cold neutrons. From ref.\ \cite{Julong} we expect,
that one can obtain an optical potential value of cC$_3$N$_4$ as high as $V \approx$ 391 neV
($b$(C) = 6.646 fm, $\rho$(cC$_3$N$_4 \approx $ 4 g/cm$^3$ \cite{sears}) with 
$\eta$ being similar to the isotopically pure cubic boron nitride.
For a coating with boron carbide, $^{11}$B$_4$C, an very low loss parameter $\eta$
is expected, $\eta \approx 10^{-7}$ ($\sigma_\mathrm{a}$ = 0.0035 barn) with the optical potential
$V \approx$ 235 neV ($\rho$ = 2.52 g/cm$^3$).

\noindent
{\bf \large Acknowledgements}\\
We thank the crew of the reactor TRIGA Mainz, J.\ Breuel,
H.\,O.\ Kling, A.\ Schmidt, and H.-M.\ Schmidt for their help
during the experiments.
The experiment benefitted from the help of M.\ Meier, PSI, in building the chopper.
This work would not have been possible without help from E.\ Gries, who supplied the
liquid helium for our superthermal UCN source.
This work was supported by
the DFG under the contract numbers KR 1458/8-2 and HE 2308/2-3.

\newpage
\begin{table}[h]
\begin{center}
\begin{tabular} {|l|c|c|c|c|}
\hline
\multicolumn{5}{|l|}{Transmission experiment}\\
\hline
Sample &\quad V(Eq.(2))[neV]&\quad V(exp)[neV]&\quad v$_c$(exp)[m/s]&\quad $\Omega$[Ohm$\cdot $cm]\\
 \hline
Ni(spin up) &\quad 289&\quad (284$\pm$15)&\quad (7.37$\pm$0.19)&\quad --\\
 \hline
Ni(spin down)&\quad 211&\quad (207$\pm$15)&\quad (6.29$\pm$0.23)&\quad --\\
 \hline
cBN$^*$)&\quad 308 &\quad (305$\pm$15)&\quad (7.64$\pm$0.19)&\quad > $3.4\cdot 10^{16}$\\
 \hline
diamond &\quad 305&\quad - -&\quad - - &\quad $3.2\cdot 10^{16}$\\ 
 \hline
\multicolumn{5}{|l|}{Cold neutron reflectometry}\\
\hline
cBN$^*$) &\quad 308 &\quad (300$\pm$30)&\quad (7.58$\pm$0.38)&\quad --\\
\hline
\end{tabular}
\end{center}
$^*$)90$\%$ cubic phase
\caption{Measured wall-potentials $V$ of natural Ni, cBN, diamond and resulting
critical velocities $v_c$ together with the measured specific resistivities for cBN and diamond.
In order to give more confidence on our analysing procedure, a Ni
layer was measured additionally to verify $V_{\rm Ni}$ as expected from literature.
The resulting error on $V$ (30 neV) measured using cold neutron reflectometry
mainly comes from the uncertainty
to give an absolute value on the angle of grazing incidence.}
\label{results}
\end{table}

\newpage
\noindent
{\bf \large Figure captions:}\\
Fig.\ \ref{fig_chopper}:\\
Sketch of the experimental setup for the time-of-flight experiment with
a chopper.

\noindent
Fig.\ \ref{fig_org}:\\
Measured time-of-flight data (i) without sample (quads),
(ii) with the nickel coated silicon wafer (triangles) and the cBN coated one
(circles).

\noindent
Fig.\ \ref{fig_cor}:\\
Measured time-of-flight data after background subtraction
(i) without sample (quads),
(ii) with the nickel coated silicon wafer (triangles) and the cBN coated one
(circles).

\noindent
Fig.\ \ref{fig_trapez}:\\
Opening function of the chopper measured at a frequency of 0.1 Hz.
In the analysis, the opening function is approximated by the
trapezium as indicated.

\noindent
Fig.\ \ref{fig_5}:\\
Time-of-flight data of very slow neutrons
through coated silicon wafers and the fit
to the data.
Top: 500 nm nickel coated on a 380 $\mu$m silicon wafer.
Bottom: 350 nm cBN on a 380 $\mu$m silicon wafer.
The corresponding TOF-data without sample
used to calculate the spline fitting function for unfolding the data
with the time acceptance of the chopper are shown in both plots.
The measured data
were corrected for the chopper time offset $\delta$t (see text).

\noindent
Fig.\ \ref{fig_6}:\\
Calculated transmission of the measured silicon wafers coated with Ni (500 nm) and cBN (300 nm) layers.
The calculation was performed by using the obtained fit parameters.
The theoretical transmission of an isotopically enriched c$^{11}$BN (300nm) layer was calculated
from literature cross sections. The saw-shape behaviour of transmission curves in some
places is the result of quantum interference at thin layers.

\noindent
Fig.\ \ref{fig_7}:\\
Measured cold neutron reflectometry data from a silicon wafer coated with 300 nm of
cBN and the corresponding fit.

\newpage

\newpage

\newpage

\newpage

\newpage

\newpage

\newpage

\end{document}